\documentclass{article}
\usepackage{authblk}
\usepackage{apacite}
\usepackage{natbib}
\usepackage{url}
\usepackage{float}
\usepackage{booktabs}
\usepackage{colortbl}
\usepackage{graphicx}

\title{Astronomy in Colombia: a bibliometric perspective}

\author[1]{\textbf{Sof\'ia Guevara-Montoya}}
\author[2]{\textbf{Felipe Ortiz-Ferreira}}
\author[1]{\textbf{María Paula Silva-Arevalo}}
\author[3]{\textbf{ Paola A. Niño-Muñoz}}
\author[3,4]{\textbf{Jaime E. Forero-Romero}}

\affil[1]{Departamento de Física, Universidad Nacional de Colombia - Sede Bogotá, Av. Cra 30 No. 45-03 - Edificio Uriel Gutiérrez, CP 111321, Bogotá, Colombia.}
\affil[2]{Escuela de Geología, Universidad Industrial de Santander, Cra. 27 No. 9, Ciudad Universitaria, CP 680002, Bucaramanga, Colombia}
\affil[3]{Departamento de Física, Universidad de los Andes, Cra. 1 No. 18A-10, Edificio Ip, CP 111711, Bogotá, Colombia}
\affil[4]{Observatorio Astronómico, Universidad de los Andes, Cra. 1 No. 18A-10, Edificio H, CP 111711, Bogotá, Colombia}

\begin{document}

\maketitle
\section*{Resumen}
En Colombia, la investigación astronómica está experimentando un crecimiento acelerado. Para comprender mejor su evolución y estado actual, realizamos un estudio bibliométrico utilizando el Astrophysics Data System (ADS) y la Web of Science (WoS).

En el ADS, identificamos 422 publicaciones arbitradas desde 1980, año de la primera publicación, hasta 2023, año de corte del estudio.
De las 25 instituciones colombianas participantes en al menos una publicación, 14 son de origen privado y 11 son instituciones estatales.
Más de la mitad de estas instituciones se concentran en dos ciudades principales: Bogotá con 11 instituciones, seguida por Medellín con 3 instituciones.

Destacan el número de contribuciones de cuatro universidades: la Universidad de los Andes, la Universidad Nacional de Colombia, la Universidad Industrial de Santander y la Universidad de Antioquia con 104, 78, 68 y 67 publicaciones, respectivamente. 
Al cruzar la información del ADS y la WoS, encontramos que las áreas en las que se encuentran las publicaciones de mayor impacto son tres: altas energías y física fundamental, estrellas y física estelar, y galaxias y cosmología.

A nivel global, según la WoS, Colombia se encuentra en el puesto 52 en cantidad de publicaciones arbitradas entre 2019 y 2023, y en quinto lugar en América Latina.
Además, identificamos tres publicaciones altamente citadas (top 1\% mundial) pertenecientes al área de cosmología observacional. 
\\[0.2cm]
\textbf{Palabras clave:}    Investigación astronómica; Análisis bibliométrico; Análisis de citaciones; Instituciones colombianas; Rankings globales

\section*{Abstract}

In Colombia, astronomical research is experiencing accelerated growth. To better understand its evolution and current state, we conducted a bibliometric study using data from the Astrophysics Data System (ADS) and the Web of Science (WoS).

In the ADS, we identified 422 peer-reviewed publications from 1980, the year of the first publication, until 2023, the cut-off year of the study.
Of the 25 Colombian institutions participating in at least one publication, 14 are private and 11 are state institutions.
More than half of these institutions are concentrated in two main cities: Bogotá with 11 institutions, followed by Medellín with 3 institutions.

The number of contributions from four universities stands out: Universidad de los Andes, Universidad Nacional de Colombia, Universidad Industrial de Santander, and Universidad de Antioquia with 104, 78, 68, and 67 publications, respectively.
By cross-referencing the information from the ADS and the WoS, we found that the areas in which publications with the highest impact are found are three: high energies and fundamental physics, stars and stellar physics, and galaxies and cosmology.

At the global level, according to the WoS, Colombia ranks 52nd in the number of peer-reviewed publications between 2019 and 2023 and fifth in Latin America.
Additionally, we identified three highly cited publications (top 1\% worldwide) belonging to the field of observational cosmology. \\[0.2cm]
\textbf{Keywords:} Astronomical research; Bibliometric analysis; Citation analysis; Colombian institutions; Global rankings

\section*{Introducción}

La astronomía desempeña un papel protagónico en la investigación científica a nivel mundial, un rol que se ha consolidado durante cerca de tres milenios, como lo demuestran los primeros registros de observaciones astronómicas sistemáticas  \citep{evans1998history}.
Desde la revolución científica iniciada por Galileo en el siglo XVII \citep{Galileo}, la astronomía no solo nos proporciona una comprensión profunda del universo y sus complejidades, sino que también contribuye significativamente al desarrollo tecnológico, generando, por ejemplo, tecnologías para la creación de imágenes con diversas aplicaciones en agricultura, navegación y salud \citep{Impacto}. 

En el último siglo, la astronomía también ha impulsado la exploración espacial a través de misiones espaciales que recopilan datos cruciales y realizan descubrimientos sobre nuestro sistema solar y más allá \citep{Exploracion}.
Actualmente, el estudio de la evolución del universo, desde la formación de estrellas hasta la existencia de la materia oscura, pasando por la formación de sistemas planetarios, permite a los científicos explorar los aspectos más fundamentales de la realidad física  \citep{materiaoscura}.  

En el siglo XX, la investigación astronómica experimentó una revolución con el desarrollo de tecnologías como el telescopio espacial y el uso de detectores más sensibles \citep{Telescopios}.
El avance en la comprensión de la física cuántica y la teoría de la relatividad influyó significativamente en la interpretación de datos astronómicos \citep{Cuantica}.
En el siglo XXI, la investigación astronómica ha entrado en la era del big data, con telescopios avanzados generando cantidades masivas de información \citep{Bigdata}.
El procesamiento computacional y el análisis de datos se han vuelto críticos, requiriendo una convergencia de habilidades en astronomía y ciencias de la computación \citep{computacional}.
Además, la cooperación global se ha intensificado aún más, con proyectos que involucran colaboraciones internacionales \citep{Colaboracion}.
En conjunto, la astronomía no solo expande nuestro conocimiento cósmico, sino que también cataliza avances tecnológicos, científicos, educativos y diplomáticos a escala global \citep{Tecnologia}.

En este quehacer, las publicaciones arbitradas desempeñan un papel crucial en la investigación astronómica al ser la principal vía para comunicar y validar los resultados científicos \citep{resultados}. 
Estas revistas científicas, revisadas por pares, actúan como foros donde los astrónomos comparten sus descubrimientos, teorías y observaciones con la comunidad científica global. 
El proceso de revisión por pares asegura la calidad y validez de la investigación antes de su publicación, contribuyendo a mantener altos estándares académicos \citep{peer}.

¿Cuál es el grado de inserción de Colombia en estas tendencias globales de investigación en astronomía? En este artículo, queremos empezar a dar respuesta a esta pregunta. 
Nos proponemos dejar un registro histórico sobre la investigación astronómica en Colombia, entendida como la actividad científica que sigue los estándares establecidos en el Siglo XX de recolección de datos de interés astronómico (ya sean observados o simulados), para proceder con su análisis estadístico y, a veces, una interpretación en términos de otras ciencias naturales, para luego presentar sus principales procedimientos y conclusiones a través de publicaciones arbitradas \citep{libro}.

En cuanto a los esfuerzos pasados, ya publicados por presentar una historia de la investigación astronómica, que podrían haberse informado a partir de los resultados existentes en publicaciones arbitradas, tenemos tres precedentes escritos por personas afiliadas al Observatorio Astronómico Nacional (OAN) en Bogotá. 

El primero, en 1993, cuando el historiador Jorge Arias de Greiff en su libro \emph{La Astronomía en Colombia}, ofrece una revisión histórica de la actividad astronómica en el territorio hoy conocido como Colombia desde la época precolombina \citep{Greiff}. 
Sin embargo, no hay mención a todas las actividades que permitieron la creación de las publicaciones arbitradas que ya existían en esa época. Solamente se dedican 3 páginas a hacer comentarios generales sobre las actividades del OAN desde 1985 hasta 1992, sin mencionar los avances con los que ya contaban en ese momento en la Universidad de los Andes (Uniandes).

El segundo esfuerzo se llevó a cabo en 2006, cuando el astrónomo William E. Cepeda-Peña presentó una contribución al simposio \emph{Astronomy for the Developing World} \citep{Cepeda}, ofreciendo una visión de la investigación en astronomía en Colombia que, nuevamente, se limitaba a presentar el estado del área en el Observatorio Astronómico Nacional (OAN), sin mencionar las actividades y publicaciones que ya existían hasta ese momento en otras cuatro instituciones.

El astrónomo Mario A. Higuera-Garzón presentó la contribución más reciente en 2016 durante la XV Reunión Regional Latinoamericana de la Unión Astronómica Internacional (LARIM) en Cartagena, Colombia, donde abordó el estado de la astronomía en el país \citep{2017RMxAC..49....3H}. 
En este documento, se ofrece un repaso del papel histórico del Observatorio Astronómico Nacional (OAN), seguido de una lista de grupos asociados a universidades que se dedican a la investigación, enseñanza y divulgación de la astronomía, mencionando de manera general la labor realizada en el OAN, Uniandes, UdeA y UIS. 
A pesar de ofrecer una visión más amplia, el panorama presentado sigue siendo incompleto. 
Como veremos más adelante, para ese entonces ya existían al menos otras 10 instituciones con publicaciones arbitradas en el campo de la astronomía.

Esto quiere decir que a la fecha todavía hace falta una visión integral que permita evaluar el alcance y el impacto de las investigaciones astronómicas realizadas en el país que se pueden considerar aportes a la comunidad internacional. 
En este artículo nos proponemos dar un primer paso para construir esta visión integral. 
Para eso nos basamos en analizar publicaciones arbitradas en el ámbito de la astronomía que cuentan con participación de instituciones colombianas. Las publicaciones arbitradas, además de ser fundamentales para la difusión del conocimiento y la construcción del corpus científico en astronomía, permiten evaluar la reputación de un grupo o comunidad astronómica y su impacto en la comunidad científica global, lo que subraya la importancia de este tipo de comunicaciones para la carrera y el reconocimiento profesional en la investigación astronómica \citep{2018EPJWC.18605002C,2020OAst...29..251W,2014SPIE.9149E..0AC}.

Las cifras y estadísticas que aquí presentamos, además de ser obtenidas con una metodología que reduce los sesgos de anteriores estudios para reportar la actividad en investigación astronómica hecha en Colombia, nos ofrecen la ventaja de permitirnos comparar a Colombia con otros países del mundo.

Según lo anterior, con el propósito de modelar la producción científica de astronomía en Colombia, presentamos un análisis bibliométrico. En primer lugar, abordamos la importancia y las ventajas de los diferentes criterios utilizados, como la cantidad de autores, citaciones, publicaciones, temáticas principales y la selección de países para comparaciones. 
Luego, describimos las búsquedas de información realizadas en las bases de datos empleadas (Astrophysics Data System y Web of Science). 
Tras explicar la metodología utilizada, presentamos los resultados en diversas secciones: número de publicaciones, número de citaciones, las publicaciones más citadas a nivel nacional, la relación entre el número de citaciones y el número de autores, y la posición de Colombia en el contexto mundial. 
Finalmente, cerramos con nuestras conclusiones y visión a futuro.

\section*{Aspectos principales de nuestro análisis bibliométrico}

\subsection*{Énfasis en publicaciones arbitradas}
En este estudio, nos enfocamos principalmente en el análisis del número de publicaciones arbitradas y citaciones, excluyendo consideraciones relacionadas de manera directa con la enseñanza o la divulgación, a menos que derivaran en publicaciones arbitradas. 
Por ejemplo, \textit{The Educational and Influential Power of the Sun} \citep{Calvo} es una publicación en el área de la educación y la divulgación que fue arbitrada, por lo que la incluimos en nuestro estudio. Pero no tenemos en cuenta aspectos como la organización de eventos, trabajos de pregrado y posgrado, ni cualquier otro tipo de publicación no arbitrada. 

La principal motivación de este enfoque radica en que los artículos arbitrados son ampliamente reconocidos como el principal resultado del proceso de creación de conocimiento científico. 
Otra importante motivación es la existencia de datos consolidados y ampliamente reconocidos para publicaciones arbitradas, lo que nos permite reducir sesgos al explorar el panorama de la investigación astronómica en todo el país.
En contraste, no existen bases de datos homogeneizadas a nivel nacional e internacional para aspectos relacionados con la enseñanza. 
De igual manera, la divulgación científica carece frecuentemente de un registro escrito que pueda consultarse.

Para evitar ambigüedades en el análisis, decidimos  utilizar dos plataformas de gran reconocimiento: Astrophysics Data System (ADS) y Web of Science (WoS). Empleamos ADS principalmente para mediciones históricas de la investigación en Colombia, mientras que WoS tiene una utilidad principal para realizar comparaciones con otros países del mundo.

\subsection*{Citaciones como índice de impacto}

En este artículo utilizamos las citaciones como medida del impacto de las publicaciones de una comunidad científica. 
Esto tiene varias ventajas. En primer lugar, las citaciones nos dan una forma objetiva de evaluar la influencia y relevancia de un artículo en el campo científico, ya que representan el reconocimiento y la validación por parte de otros investigadores. 
Además, al mediar las citaciones a lo largo del tiempo, es posible analizar la evolución del impacto de un trabajo o área y su contribución continua al avance del conocimiento en un área determinada \citep{citaciones}. 
Esto nos va a permitir identificar las publicaciones más influyentes y destacadas en el área de la astronomía, así como entender las áreas de mayor productividad e impacto al interior de las principales universidades del país.

Sin embargo, se deben tener en cuenta las limitaciones importantes de este tipo de mediciones de impacto a través de citaciones. 
Por ejemplo, las citaciones pueden ser influenciadas por factores externos, como el prestigio del autor o la revista donde se publicó el artículo, lo que puede distorsionar la percepción real del impacto de un trabajo \citep{prestigio}. 
Además, el tiempo para acumular citaciones varía, lo que dificulta la comparación entre diferentes publicaciones.

En este artículo también vamos a utilizar el índice-h, o índice de Hirsch \citep{Hirsh}, para intentar abordar algunas de las limitaciones de las citaciones individuales al tener en cuenta las citaciones recibidas por un conjunto de publicaciones en periodos extensos de tiempo. 
Este índice se define como el número de artículos dentro de un conjunto que han sido citados al menos h veces cada uno. 
Si un conjunto de publicaciones tiene un índice-h de 10, significa que tiene al menos 10 artículos que han sido citados al menos 10 veces cada uno.

Finalmente, como medida del impacto de la investigación astronómica con participación colombiana en el contexto internacional, vamos a utilizar el concepto de artículo altamente citado de WoS. 
Un artículo altamente citado es aquel que se encuentra en el 1\% superior de los artículos con mayores citaciones a nivel mundial en el año de su publicación.

Seria sumamente enriquecedor para la comunidad que en futuros estudios pudiese incluirse un análisis de las instituciones y países con los que más colabora Colombia y palabras clave o palabras más comunes utilizadas en los resúmenes y títulos de las publicaciones con el fin de llevar a cabo un análisis de redes bibliométricas más complejo.

\subsection*{Número de Autores}

Otro aspecto importante a cuantificar es el número de autores. 
Esto nos permite observar la importancia de las colaboraciones en el desarrollo de artículos científicos. 
Además, nos permite identificar diferentes redes científicas que operan en Colombia y evaluar el impacto de la contribución individual y colectiva.

Aunque el número de autores no determina la calidad del artículo, cuando hay un número elevado de autores, el artículo tiende a ser más conocido, lo que incrementa su número de citaciones y lo hace más destacado \citep{factores}. 
Por eso, también cuantificamos el número de autores en las publicaciones para estimar hasta qué punto esto se correlaciona la cantidad de autores con un mayor número de citaciones.

\subsection*{Temáticas}

Las temáticas son un factor crucial al analizar los artículos, ya que representan los temas destacados de trabajo en el país y de cada institución. 
Sin embargo, nos enfrentamos a un reto, ya que, al manejar grandes bases de datos de todos los artículos, es complicado determinar el tema principal de cada uno. 
Por esta razón, solamente clasificamos los artículos de mayor impacto a partir de la clasificación de las nueve divisiones de trabajo establecidas por la Unión Astronómica Internacional (UAI):

\begin{itemize}
    \item División A: Astronomía Fundamental
    \item División B: Instalaciones, Tecnologías y Ciencia de Datos
    \item División C: Educación, Divulgación y Patrimonio
    \item División D: Fenómenos de Altas Energías y Física Fundamental
    \item División E: Sol y Heliosfera
    \item División F: Sistemas Planetarios y Astrobiología
    \item División G: Estrellas y Física Estelar
    \item División H: Materia Interestelar y Universo Local
    \item División J: Galaxias y Cosmología
\end{itemize}

Con lo anterior, podemos determinar cuáles son las áreas de investigación de las publicaciones más influyentes en nuestro país, para revisar las fortalezas que se tienen en diferentes campos de estudio e identificar cuáles temas podrían presentar menos interés o enfrentar problemáticas que no permiten su desarrollo al mismo nivel.

\subsection*{Selección de países para comparativos}

Nos interesa comparar la actividad astronómica colombiana con la de otros países del mundo. 
Para ello, hemos elegido comparar el número total de publicaciones en un período reciente de cinco años, del 2019 al 2023, utilizando los datos de WoS.

Los países que incluimos en nuestra comparación se dividen en dos grupos. El primero comprende aquellos países que tienen un número igual o mayor de publicaciones que Colombia. 
El segundo grupo está conformado por países de América Latina y el Caribe que tengan al menos una publicación. 
Esto nos permite compararnos con naciones que comparten similitudes lingüísticas, culturales e históricas, así como desafíos comunes en el desarrollo científico y tecnológico.

\section*{Bases de datos}

Para nuestro análisis, utilizamos dos bases de datos diferentes: el Astrophysics Data System (ADS) y Web of Science (WoS).

El Astrophysics Data System (ADS) es un sistema de búsqueda y recuperación de información académica en el campo de la astronomía y la astrofísica. Ampliamente utilizado en la comunidad astronómica, es una herramienta valiosa para la investigación y el acceso a la información en este campo. 
La plataforma es mantenida por la National Aeronautics and Space Administration (NASA) de los Estados Unidos y su interfaz web es de acceso público.

Web of Science (WoS) es una plataforma en línea que proporciona acceso a una amplia gama de recursos académicos y científicos. Desarrollada por Clarivate Analytics, es conocida por ser una de las bases de datos más completas y respetadas para la investigación científica y académica. El acceso completo a sus recursos requiere una suscripción paga.

Dentro de ambas bases de datos utilizaremos la clasificación temática interna más cercana a la astronomía.
En el caso de ADS, los artículos pueden recibir cuatro clasificaciones de temáticas diferentes: Astronomía, Física, Ciencias de la Tierra y General.
En WoS, la clasificación abarca 10 macrotemas. Astronomía y Astrofísica es un mesotema específico que se encuentra bajo el macrotema Física. Otros mesotemas relacionados son, por ejemplo, Partículas y Campos, Ciencias Meteorológicas y Atmosféricas, y Ciencias Espaciales.

Para enfocarnos en la productividad de la investigación que sin duda es reconocida como astronomía, solamente incluimos las publicaciones que se clasifican \emph{exclusivamente} en el área "Astronomy" (en ADS) o "Astronomy \& Astrophysics" en WoS. 
Esto tiene el efecto principal de excluir publicaciones que se encuentran dentro de la física de partículas, la teoría de campos y la gravitación.
Además, en el caso de ADS, seleccionaremos explícitamente registros de 
publicaciones arbitradas para evitar la inclusión de publicaciones no revisadas por pares del repositorio arXiv, las cuales ya se encuentran excluidas de los resultados de WoS. 

Los resultados de ADS solamente los utilizamos para obtener estadísticas sobre Colombia, pero no sobre el resto de países.
La gran ventaja de ADS sobre WoS es que incluye artículos desde 1800, lo cual nos sirve para trazar una historia sobre las primeras épocas de publicaciones arbitradas.

\subsection*{Astrophysics Data System}

El query que establecimos para la página de Astrophysics Data System (ADS) realiza una búsqueda de todos los artículos en los que la afiliación de los autores contenga la palabra "Colombia". 
A diferencia de WoS, ADS no indexa las publicaciones con el país de origen de las instituciones asociadas a los autores. 
Por lo tanto, nos vemos obligados a utilizar la palabra "Colombia" en la búsqueda, lo que conlleva el riesgo de encontrar publicaciones que no estén directamente relacionadas con el país, sino que simplemente incluyan una institución con la palabra "Colombia". 
Para mitigar este riesgo, hemos intentado eliminar las combinaciones de palabras que encontramos podrían generar confusiones, como "Colombia Astrophysics Laboratory", "British Colombia" o "Av. Gran Colombia".
Los resultados de esta búsqueda están disponibles en la siguiente página web de ADS: \textbf{\url{https://ui.adsabs.harvard.edu/public-libraries/H5bjkAlXSYaZcO4qGcKxOw}}.

Además, hemos realizado 4 búsquedas adicionales para filtrar datos de las cuatro instituciones que más publican en el país, según la misma clasificación que realiza ADS. 
Estas instituciones son: la Universidad de los Andes, Universidad Nacional de Colombia, la Universidad Industrial de Santander y la Universidad de Antioquia.

Otros resultados que obtenemos de ADS incluyen:
\begin{itemize}
\item Las citaciones por cada artículo.
\item El número de autores para cada artículo.
\item El cálculo del índice-h para grupos de artículos.
\end{itemize}

Las cifras que utilizamos de ADS corresponden a marzo del 2024.

\subsection*{Web of Science }

En Web of Science (WoS), realizamos un query buscando en las "Web of Science Categories" todas las entradas bibliográficas clasificadas como "Astronomy \& Astrophysics". 
Luego, dentro de la categoría "Citation Topics Meso", nos enfocamos exclusivamente en las entradas clasificadas dentro de "5.20 Astronomy and Astrophysics" hasta el año 2023.
Finalmente, seleccionamos los artículos que WoS indexa como asociados a Colombia.
Para asegurarnos de tener un conjunto de datos representativo de la actividad reciente, creamos otro conjunto de datos donde restringimos las publicaciones a los años 2019 hasta 2023, inclusive. 

Es importante destacar que los datos que utilizamos corresponden a un query realizado en enero de 2024, cuando WoS aún incluía publicaciones desde 1900. Sin embargo, la versión actual de WoS (posterior al 4 de febrero de 2024) solamente considera publicaciones en esta área realizadas desde 1988 hasta la fecha.

A partir de los resultados obtenidos en WoS, vamos a utilizar los siguientes datos calculados dentro de la misma base de datos:
\begin{itemize}
    \item La evolución temporal de las publicaciones y citaciones para las publicaciones realizadas en Colombia.
    \item El número de citaciones de las publicaciones más citadas de las principales universidades de Colombia ya identificadas.
    \item El número total de publicaciones y las publicaciones altamente citadas para los países de referencia en el periodo de cinco años (2019-2023).
\end{itemize}

\section*{Resultados}

\subsection*{Número de publicaciones}

En el ADS, encontramos un total de 422 publicaciones arbitradas relacionadas con Colombia. 
La Tabla \ref{tab:universidades} presenta información detallada sobre las 25 instituciones que logramos identificar como participantes en al menos una de estas publicaciones. 
En esta tabla, se muestran el índice-h, el número total de publicaciones, el número total de citas para cada institución y el año de la primera publicación de esa institución dentro del total de las publicaciones encontradas.
Destacan cuatro instituciones con la mayor cantidad de artículos: la Universidad de los Andes, la Universidad Nacional de Colombia, la Universidad Industrial de Santander y la Universidad de Antioquia, con 104, 78, 68 y 67 artículos respectivamente.

A partir de esta tabla es posible ver que para la fecha de la publicación del libro La astronomía en Colombia \citep{Greiff}, ya había dos universidades con publicaciones arbitradas en el área de astronomía.
En el 2005, al momento de la presentación en el simposio \emph{Astronomy for the developing world} \citep{Cepeda}, eran cinco las instituciones que contaban con publicaciones.
Mientras que en el 2016, al momento de la presentación en la XV LARIM con el tema \emph{Astronomy in Colombia} \citep{2017RMxAC..49....3H}, eran 15 las universidades con participación en actividades de investigación que quedaron registradas en forma de publicaciones arbitradas.

Según ADS, los primeros artículos se remontan a 1980. Uno de ellos, publicado en la Revista Colombiana de Física por José Granés, un físico colombiano y profesor del Departamento de Física de la Sede Bogotá de la Universidad Nacional de Colombia, aborda una comparación de los conceptos de espacio y tiempo en física clásica y la relatividad especial.

Este texto de carácter expositivo no ha recibido citaciones hasta la fecha \citep{1980RvCF...14..115G}.
El otro artículo, publicado en agosto de 1980, está escrito por un astrónomo alemán y profesor del Departamento de Física de la Universidad de los Andes en Bogotá.
Este artículo, que se encuentra en las \emph{Publications of the Astronomical Society of the Pacific}, presenta un análisis de datos fotométricos de la estrella cefeida AH Velorum para sugerir que esta estrella es en realidad binaria y tiene una compañera.
Este texto investigativo ha recibido seis citaciones hasta la fecha \citep{1980PASP...92..484G}.
Hasta donde sabemos, es Gieren la primera persona con un título de doctorado en dedicarse a la investigación astronómica en Colombia.

Un análisis de los temas y enfoques metodológicos abordados en estos dos artículos permite situar los inicios de la investigación astronómica contemporánea en Colombia en el trabajo precursor llevado a cabo por Gieren en el Departamento de Física de la Universidad de los Andes, en Bogotá.

\begin{table}[H]
\caption{Instituciones colombianas que participan en artículos de astronomía según la selección hecha en ADS. Las segunda, tercera y cuarta columnas indican el índice-h, el número de publicaciones, el número total de citaciones y el año correspondiente a la primera publicación, respectivamente. }
\label{tab:universidades}
\begin{tabular}{lp{4.2cm}cccc} 
\toprule
\textbf{Rango} & \textbf{Universidad} & \textbf{Indice-h} & \textbf{\# Pub.} & \textbf{\# Cit.} & \textbf{Año inicial} \\ \hline
\rowcolor[rgb]{0.753,0.753,0.753}
1 & Universidad de los Andes&	26	&104	&2617 & 1980\\
2 &Universidad Nacional de Colombia&	20&	78	&1010 & 1980\\
\rowcolor[rgb]{0.753,0.753,0.753}
3 &Universidad Industrial de Santander&	19	&68	&4734 & 2006\\
4 &Universidad de Antioquia &	18	&67	& 1387 & 2010\\
\rowcolor[rgb]{0.753,0.753,0.753}
5 &Universidad de Medellín &	13&	19	&534 & 2010\\
6 &Universidad del Valle&	7&	15	 &117 & 2012\\
\rowcolor[rgb]{0.753,0.753,0.753}
7 &Universidad de Nariño &	6	&10	&137 & 2015\\
8 &Universidad Tecnológica de Bolivar&	5	&10	&144 & 2011\\
\rowcolor[rgb]{0.753,0.753,0.753}
9 &Fundación Universitaria Konrad Lorenz&	5 &	7	&387 & 2016\\
10 &Universidad Tecnológica de Pereira&	4&	13	&62 & 1997\\
\rowcolor[rgb]{0.753,0.753,0.753}
11 &Universidad ECCI&	4&	8 &	50 & 2017\\
12 &Universidad de los Llanos&	4	&8	&38 & 2017\\
\rowcolor[rgb]{0.753,0.753,0.753}
13 & Centro Internacional de Física &	4&	7 &	48 & 1995\\
14 & Universidad Antonio Nariño &	4&	5 &	126 & 2006\\
\rowcolor[rgb]{0.753,0.753,0.753}
15 &Universidad Distrital Francisco José de Caldas&	2&	5&	23 & 2012\\
16 &Universidad Sergio Arboleda &	2&	3	&46 & 2010\\
\rowcolor[rgb]{0.753,0.753,0.753}
17 &Universidad Militar Nueva Granada&	2&	3	&28 & 2017\\
18 &Universidad del Atlántico &	1	&5 &	6 & 1997\\
\rowcolor[rgb]{0.753,0.753,0.753}
19 &Pontificia Universidad Javeriana& 1	&3&	5 & 2018\\
20 & Universidad EAN &	1&	2&	2 & 2023\\
\rowcolor[rgb]{0.753,0.753,0.753}
21 &Universidad El Bosque &	1	&1	& 152 & 2023\\
22 &Universidad EAFIT &	1&	1	& 8 & 2019\\
\rowcolor[rgb]{0.753,0.753,0.753}
23 &Universidad de Córdoba&1	&1	&2 & 2022\\
24 &Corporación Universitaria Autónoma del Cauca &	1&	1	&1 & 2019\\
\rowcolor[rgb]{0.753,0.753,0.753}
25 & Universidad Mariana&	0&	1	&0 & 2023\\
\bottomrule
\end{tabular}
\end{table}

\begin{figure}[H]
\centering
   \includegraphics[width=0.9\linewidth]{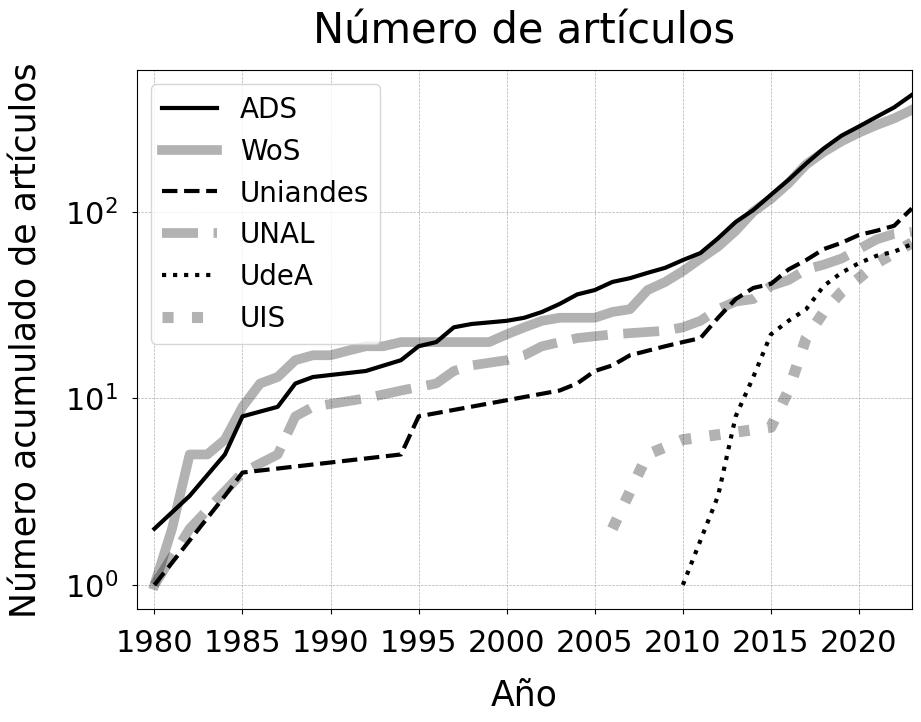}
    \caption{
Número acumulado de artículos publicados en astronomía con participación de instituciones colombianas desde la primera publicación en 1980 hasta 2023, según los datos de las bases de datos ADS y WoS. 
Se observa una tendencia de crecimiento exponencial en las publicaciones a lo largo de los años. 
Las instituciones con mayor número de publicaciones son la Universidad de los Andes (Uniandes), la Universidad Nacional de Colombia (UNAL), la Universidad de Antioquia (UdeA) y la Universidad Industrial de Santander (UIS). 
De acuerdo a los cambios presentados en la gráfica establecemos tres etapas diferentes: La fase pionera de 1980 a 1989, la fase de desarrollo de 1990 a 2009 y la fase de consolidación de 2010 a 2019. }
    \label{fig:papers_acumulados}
\end{figure}

La figura \ref{fig:papers_acumulados} muestra el número acumulado de artículos publicados hasta un año determinado, desde 1980 hasta 2023. 
Las dos líneas continuas representan los resultados de ADS y WoS, mientras que las otras líneas muestran los resultados de las cuatro universidades con mayor producción.

Definimos tres fases distintas a partir de los cambios presentados en la gráfica para las publicaciones totales de Colombia. 
En la primera fase, que abarca aproximadamente una década desde 1980 hasta 1989, se supera apenas la primera decena de publicaciones a lo largo de la década.
A esta década la denominamos la fase pionera.

En la segunda fase, que abarca alrededor de dos décadas desde 1990 hasta 2009, se publican cerca de 40 artículos en total, con una tasa promedio de alrededor de 20 artículos por década, multiplicando por dos la tasa de publicaciones de la fase anterior.
Durante estas dos primeras fases, se observa una clara hegemonía de dos universidades en Bogotá: la Universidad Nacional en primer lugar y la Universidad de los Andes en segundo lugar.
A estas dos décadas las denominamos la fase del desarrollo. 

Desde 2010 hasta 2019, entramos en una tercera fase en la que se registran cerca de 200 artículos nuevos, multiplicando por diez la tasa de publicaciones por década en comparación con la fase anterior. 
Además, la hegemonía de Bogotá se comparte ahora con las principales universidades de Medellín y Bucaramanga: la Universidad de Antioquia y la Universidad Industrial de Santander. 
En esta década, la Universidad de los Andes se convierte en la institución con el mayor número de publicaciones en total. 
A esta década la denominamos fase de consolidación. 

Finalmente, iniciando una nueva década, de 2020 a 2023, se registran cerca de 170 publicaciones nuevas, lo que sugiere que hacia 2029 se habrán publicado alrededor de 400 artículos nuevos, duplicando la tasa de publicaciones de la década anterior.

\subsection*{Número de citaciones}

\begin{figure}[H]
\centering
    \includegraphics[width=0.9\linewidth]{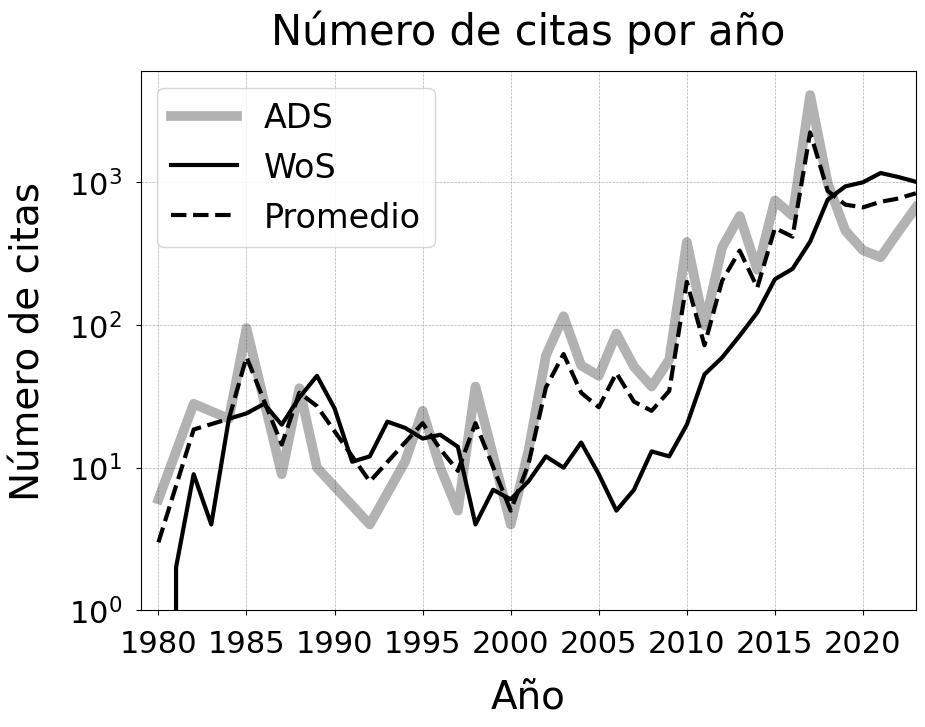}
     \caption{
    Número de citaciones recibidas por año para las publicaciones astronómicas con participación de instituciones colombianas desde 1980 hasta 2023, según los datos de las bases de datos ADS y WoS, así como su respectivo promedio. 
    Se observa una tendencia creciente en el número de citas a lo largo del tiempo, con fluctuaciones año tras año. 
    A partir de aproximadamente 2005, hay un aumento más pronunciado en el número de citas recibidas anualmente, lo que sugiere un mayor impacto de las investigaciones colombianas en astronomía en los últimos años. Las citas anuales apenas superan las 10 citaciones por año en la fase pionera (1980-1989). 
    Durante la fase de desarrollo (1990-2009) incrementa a 100 citaciones por año aproximadamente. 
    Finalmente, hacia el cierre de la fase de consolidación (2010-2019) se llega a cerca de 1000 citaciones anuales. }
     \label{fig:citaciones}
\end{figure}

En la Figura \ref{fig:citaciones} se presenta el número de citaciones por año a los artículos publicados hasta la fecha indicada. 
Las dos líneas punteadas muestran los resultados obtenidos de ADS y WoS, mientras que la línea continua representa el promedio entre ambas fuentes. Este promedio nos permite obtener una tendencia más robusta  ya que se tienen en cuenta los resultados en ambas bases de datos. 
Estos resultados reflejan una historia similar a la observada en las diferentes fases separadas por décadas que se identificaron en la sección anterior basada en el número de publicaciones. 

Durante la primera década, desde 1980 hasta 1989, la tasa de citaciones anuales apenas supera las 10 citaciones por año. 
En la siguientes dos décadas, de 1990 a 2009, se observa un incremento hasta aproximadamente 100 citaciones por año, y este aumento continúa hasta la década de 2010 a 2019, donde la tasa de citaciones crece casi diez veces, alcanzando cerca de 1000 citaciones anuales. 

Las 422 publicaciones colombianas en ADS acumulan un número total de 10946 citaciones, correspondiente a un promedio de 25 citaciones por artículo.
Las citaciones acumuladas por las cuatro universidades líder son: 2617 (Uniandes), 1010 (UNAL), 4734 (UIS) y 1387 (UdeA).
Ademas, el índice-h de las publicaciones colombianas como un todo es de 42, mientras que Uniandes, UNAL, UIS y UdeA tienen índices-h de 26, 20, 19 y 18, respectivamente.

\subsection*{Las publicaciones más citadas a nivel nacional}

Para ampliar aún más el panorama de las publicaciones e instituciones con mayor impacto, evaluamos cuáles son los artículos con el mayor número de citaciones. 
Estas citaciones se calculan a partir del promedio entre ADS y WoS. 
Por esta razón, las publicaciones que consideramos en esta sección deben encontrarse en ambas bases de datos.

La Tabla \ref{tab:top10_completo} resume la información respecto a los 10 artículos más citados, que representan aproximadamente el top 2\% de las publicaciones con participación colombiana. 
Incluimos el título, el año de publicación, el número de citaciones, el número de autores, las universidades colombianas involucradas y la clasificación UAI.

A partir de esta tabla, podemos extraer varias observaciones interesantes:
\begin{itemize}
\item Tres áreas de conocimiento dominan esta tabla: Fenómenos de altas energías y física fundamental, estrellas y física estelar, galaxias y cosmología.
\item Cada una de estas tres áreas corresponde a tres universidades: UIS para altas energías, UdeA para física estelar y Uniandes para galaxias y cosmología.
\item La UNAL, a pesar de tener un número relativamente alto de publicaciones, no logra posicionar ninguna de sus publicaciones en el Top 10 de citaciones.
\item Las publicaciones con participación de la UIS tienen cifras de coautores con varios órdenes de magnitud mayores que las otras publicaciones, lo cual se explica por su participación en el Observatorio de Rayos Cósmicos Pierre Auger.
\item Todos los artículos más citados fueron publicados en la década de 2010 a 2019.
\end{itemize}

Para ampliar aún más la visión de las publicaciones más citadas, ahora listamos el Top 5 de publicaciones más citadas para cada una de las cuatro universidades líderes. 
La misma información que se presenta en la Tabla \ref{tab:top10_completo} se proporciona para Uniandes, UNAL, UIS y UdeA en las Tablas \ref{tab:top5_andes}, \ref{tab:top5_unal}, \ref{tab:top5_uis} y \ref{tab:top5_ant}, respectivamente.

Para Uniandes, observamos que los artículos más influyentes se centran todos en el área de Galaxias y Cosmología, publicados en un período entre 2013 y 2018, durante la fase de consolidación, 2010-2019. 
Estas publicaciones implican colaboraciones pequeñas con un rango de 7 a 32 autores, y todas superan la centena de citaciones.

En el caso de la UNAL, los artículos más influyentes abarcan dos temáticas: estrellas y física estelar, galaxias y cosmología. 
Los primeros cuatro artículos implican la participación del OAN, mientras que el último corresponde a participantes del Departamento de Física. 
Estas publicaciones se distribuyen en un período que abarca desde 2002 hasta 2012, cerca del final de las décadas de desarrollo. 
Se destaca que todas estas publicaciones corresponden a un grupo pequeño de colaboradores, con cuatro autores en cada caso, y ninguna supera la centena de citaciones.

\begin{table}[H]
\caption{Top 10 de artículos con más citaciones. La segunda columna tiene el número de citaciones promedio entre ADS y WoS. La tercera columna tiene el número de autores del artículo. La cuarta columna lista las universidades colombianas que tienen participación en el artículo referenciado. La última columna presenta las división correspondiente de la UAI.}
\label{tab:top10_completo}
\begin{tabular}{p{4.5cm}cccp{2cm}} 
\toprule
\textbf{Titulo y referencia} & \textbf{Citaciones} & \textbf{Autores} & \textbf{Universidades} & \textbf{División IAU} \\ \hline
\rowcolor[rgb]{0.753,0.753,0.753} Multi-messenger Observations of a Binary Neutron Star Merger, \cite{Rango1} & 2468 & 3630 & UIS &  Fenómenos de altas energías y física fundamental \\ 
A Library of Theoretical Ultraviolet Spectra of Massive, Hot Stars for Evolutionary Synthesis, \cite{Rango2} & 253 &  8 & UdeA & Estrellas y física estelar \\
\rowcolor[rgb]{0.753,0.753,0.753} Tracing the cosmic web, \cite{Rango3} & 171,5 & 30 & Uniandes & Galaxias y cosmología\\
 Binary Neutron Star Mergers: A Jet Engine for Short Gamma-Ray Bursts , \cite{Rango4} & 162 &  4 & UIS & Fenómenos de altas energías y física fundamental \\
\rowcolor[rgb]{0.753,0.753,0.753} An Indication of Anisotropy in Arrival Directions of Ultra-high-energy Cosmic Rays through Comparison to the Flux Pattern of   Extragalactic Gamma-Ray Sources, \cite{Rango5} & 161,5 & 392 & UIS & Fenómenos de altas energías y física fundamental\\
The MultiDark Database: Release of the Bolshoi and MultiDark cosmological simulations, \cite{Rango6} & 153,5 &  11 & Uniandes & Galaxias y cosmología \\
\rowcolor[rgb]{0.753,0.753,0.753} Search for High-energy Neutrinos from Binary Neutron Star Merger GW170817 with ANTARES, IceCube, and the Pierre Auger Observatory, \cite{Rango7} & 153,5 & 1941 & UIS & Fenómenos de altas energías y física fundamental\\
(Almost) Dark Galaxies in the ALFALFA Survey: Isolated HI-bearing  Ultra-diffuse Galaxies, \cite{Rango8} & 151,5 &  12 & Uniandes & Galaxias y cosmología \\
\rowcolor[rgb]{0.753,0.753,0.753} Probing the role of the galactic environment in the formation of stellar clusters, using M83 as a test bench, \cite{Rango9} & 147 & 5 & UdeA & Estrellas y física estelar\\
The velocity shear tensor: tracer of halo alignment, \cite{Rango10} & 111,5 &  7 & Uniandes & Galaxias y cosmología \\
\bottomrule
\end{tabular}
\end{table}

Para la UIS, resalta el hecho de que cuatro de las publicaciones implican un alto número de autores, casi dos órdenes de magnitud mayor que las de Uniandes y UNAL. Todas las publicaciones se centran en el área de Altas Energías y Física Fundamental. Estas fueron realizadas entre 2016 y 2019, hacia el final de la década de consolidación. 
La mayoría de estas publicaciones supera la centena de citaciones promedio, y es notable que la publicación más citada tiene más de 2000 citaciones y más de 3500 autores.

Por último, para la UdeA, observamos que el tema dominante es estrellas y física estelar. Solo tres de estos artículos superan las 100 citaciones, y todos fueron publicados entre 2010 y 2015, al inicio de la fase de consolidación. 
Se destaca que estas publicaciones corresponden a colaboraciones más pequeñas, con un rango de autores de 4 a 24.

En un futuro trabajo seria de gran interés hacer un estudio que permita entender y cuantificar el aporte de los autores vinculados a instituciones colombianas, este estudio requeriría de un análisis más cuidadoso y detallado puesto que no siempre el orden de los autores es acorde a su aporte, algunas veces la lista de autores esta por ejemplo, en orden alfabético.

\begin{table}[H]
\caption{Top 5 de artículos con más citaciones (promedio entre ADS y WoS) publicados con participación de la Universidad de los Andes.}
\label{tab:top5_andes}
\begin{tabular}{p{ 5.5cm}ccp{3.5cm}} 
\toprule
\textbf{Titulo y referencia} & \textbf{Citaciones} & \textbf{Autores}  & \textbf{División IAU} \\ \hline
\rowcolor[rgb]{0.753,0.753,0.753} Tracing the cosmic web, \cite{Rango3} & 171,5 & 30 & Galaxias y cosmología \\ 
The MultiDark Database: Release of the Bolshoi and MultiDark cosmological simulations,\cite{Rango6} & 153,5 & 11 & Galaxias y cosmología\\ 
\rowcolor[rgb]{0.753,0.753,0.753} (Almost) Dark Galaxies in the ALFALFA Survey: Isolated HI-bearing Ultra-diffuse Galaxies, \cite{Rango8} & 151,5 & 12 & Galaxias y cosmología \\ 
The velocity shear tensor: tracer of halo alignment, \cite{Rango10}  & 113,5 & 7 & Galaxias y cosmología\\ 
\rowcolor[rgb]{0.753,0.753,0.753} The Large-scale Structure of the Halo of the Andromeda Galaxy. II. Hierarchical Structure in the Pan-Andromeda Archaeological Survey, \cite{Rank5} & 109 & 32 & Galaxias y cosmología \\ 
\bottomrule
\end{tabular}
\end{table}

\begin{table}[H]
\caption{Top 5 de artículos con más citaciones (promedio entre ADS y WoS) publicados con participación de la Universidad Nacional de Colombia.}
\label{tab:top5_unal}
\begin{tabular}{p{ 5.5cm}ccp{3.5cm}} 
\toprule
\textbf{Titulo y referencia} & \textbf{Citaciones} & \textbf{Autores}  & \textbf{División UAI} \\ \hline
\rowcolor[rgb]{0.753,0.753,0.753} Spin Evolution of Accreting Young Stars. II. Effect of Accretion-powered Stellar Winds. \cite{Rank1} & 67 & 4 &  Estrellas y
física estelar  \\ 
Spin Evolution of Accreting YoungStars I. Effect of Magnetic Star-Disk Coupling, \cite{Rank2} & 63 &  4 &  Estrellas y física estelar\\
\rowcolor[rgb]{0.753,0.753,0.753} The   Near-infrared Coronal Line Spectrum of 54 nearby Active Galactic Nuclei, \cite{rank3} & 49 &  4 &  Galaxias y cosmología  \\
Near-Infrared Coronal Lines in Narrow-Line Seyfert 1 Galaxies, \cite{Rank4} & 41,5 & 4 & Galaxias y cosmología\\ 
\rowcolor[rgb]{0.753,0.753,0.753} Dwarf   spheroidal satellites of the Milky Way from dark matter free tidal dwarf galaxy progenitors: maps of orbits, \cite{Rank6} & 36,5 &  4 &  Galaxias y cosmología \\ 
\bottomrule
\end{tabular}
\end{table}

\begin{table}[H]
\caption{Top 5 de artículos con más citaciones (promedio entre ADS y WoS) publicados con participación de la Universidad Industrial de Santander.}
\label{tab:top5_uis}
\begin{tabular}{p{ 5.5cm}ccp{3.5cm}} 
\toprule
\textbf{Titulo y referencia} & \textbf{Citaciones} & \textbf{Autores}  & \textbf{División UAI} \\ 
\rowcolor[rgb]{0.753,0.753,0.753} Multi-messenger Observations of a Binary Neutron Star Merger,  \cite{Rango1} & 2468 & 3630  &  Fenómenos de altas energías y física fundamental \\ 
Binary Neutron Star Mergers A Jet Engine for Short Gamma-Ray Bursts, \cite{Rango4} &  162 &  4 &  Fenómenos de altas energías y física fundamental\\ 
\rowcolor[rgb]{0.753,0.753,0.753} An Indication of Anisotropy in Arrival Directions of Ultra-high-energy Cosmic Rays through comparison to  the Flux Pattern of  Extragalactic  Gamma-Ray Sources, \cite{Rango5} & 161,5 & 392 &  Fenómenos de altas energías y física fundamental \\ 
Search for High-energy Neutrinos from Binary Neutron Star Merger GW17081  with ANTARES, IceCube, and the Pierre Auger Observatory, \cite{Rango7}  &  153,5 &  1941 &  Fenómenos de altas energías y física fundamental \\ 
\rowcolor[rgb]{0.753,0.753,0.753} Probing the origin of ultra-high-energy cosmic rays with neutrinos in the EeV energy range using the Pierre Auger Observatory, \cite{San5} & 87,5 &  391 &  Fenómenos de altas energías y física fundamental\\
\bottomrule
\end{tabular}
\end{table}

\begin{table}[H]
\caption{Top 5 de artículos con más citaciones (promedio entre ADS y WoS) publicados con participación de la Universidad de Antioquia.}
\label{tab:top5_ant}
\begin{tabular}{p{ 5.5cm}ccp{3.5cm}} 
\toprule
\textbf{Titulo y referencia} & \textbf{Citaciones} & \textbf{Autores}  & \textbf{División UAI} \\ 
\rowcolor[rgb]{0.753,0.753,0.753} A Library of Theoretical Ultraviolet Spectra of   Massive, Hot Stars for Evolutionary Synthesis, \cite{Rango2}  &  253 &  8 &  Estrellas y
física estelar \\
Probing the role of the galactic environment in the formation of stellar clusters, using M83 as a test bench, \cite{Rango9} &  147 & 5 &  Estrellas y
física estelar \\ 
\rowcolor[rgb]{0.753,0.753,0.753} Studying the YMC population of M83: how long clusters remain embedded, their interaction with the ISM and implications for GC formation theories, \cite{Ant3} &  108 & 7 & Estrellas y
física estelar\\
Galaxy Cluster Mass Reconstruction Project - II. Quantifying scatter and bias using contrasting mock catalogues, \cite{Ant4}  & 64,5 &  24 &  Galaxias y cosmología \\ 
\rowcolor[rgb]{0.753,0.753,0.753} No evidence for significant age spreads in young massive LMC clusters, \cite{Ant5} &  64 & 4 &  Estrellas y física estelar \\
\bottomrule
\end{tabular}
\end{table}

\subsection*{Relación entre el número de citaciones y el número de autores}

Los resultados anteriores sugieren que existen dinámicas de trabajo diferentes, donde algunas publicaciones se realizan con un pequeño número de colaboradores que no superan la decena, mientras que otros tipos de trabajos implican la colaboración de cientos o miles de personas.

\begin{figure}[H]
\centering
    \includegraphics[width=0.9\linewidth]{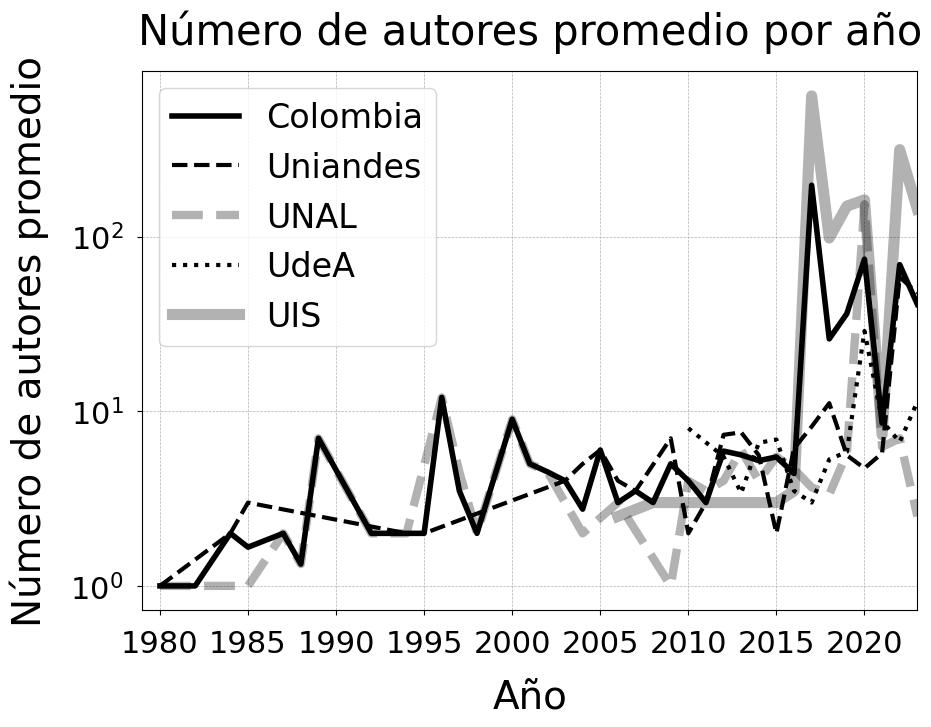}
     \caption{Promedio de número de autores por artículo en cada año. Se observa un aumento drástico de autores desde el 2016 gracias a artículos del Observatorio de Rayos Cósmicos Pierre Auger con participación de la UIS y la UdeM. Antes de eso el promedio de autores no superaba la decena.}
     \label{fig:autores_promedio}
\end{figure}

Para cuantificar la influencia de los pequeños grupos y las grandes colaboraciones, revisamos el promedio de número de autores de las publicaciones. 
La Figura \ref{fig:autores_promedio} muestra el promedio de autores desde 1980 hasta 2023. 
En esta gráfica es evidente que hasta el 2015, los artículos colombianos tienen un número promedio de autores cercano a la decena. 
Sin embargo, a partir del 2016, se observa una transición donde el número promedio de autores aumenta drásticamente, llegando a la centena. 
Este cambio se debe en gran medida a los artículos del Observatorio de Rayos Cósmicos Pierre Auger con participación de la UIS y la UdeM.

\begin{figure}[H]
    \centering
    \includegraphics[width=0.9\linewidth]{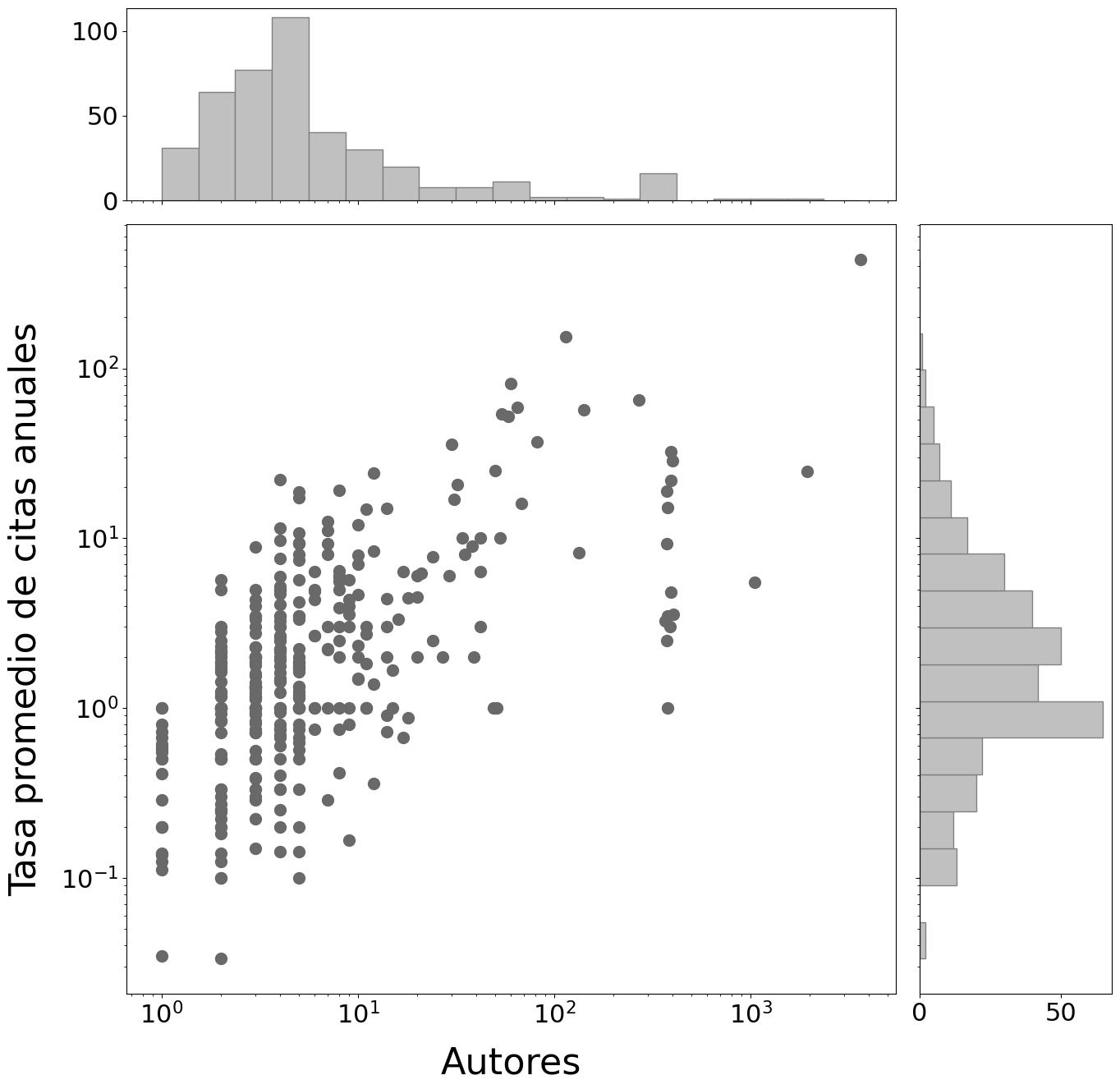}
    \caption{Tasa promedio de citaciones anuales como función del número total de autores para las publicaciones dentro de ADS que tienen al menos una citación. Estos datos presentan una correlación de Pearson de 0.78.}
    \label{fig:citation_rate}
\end{figure}

En la Figura \ref{fig:citation_rate}, exploramos la influencia del tamaño de las colaboraciones (eje horizontal, histograma de la parte superior) en la tasa anual promedio de citaciones (eje vertical, histograma del lado derecho). 
En primer lugar, observamos que la mayoría de los artículos tienen 4 o 5 autores. 
Los artículos con más de cinco autores muestran una incidencia decreciente, excepto por un exceso que se observa para los artículos con entre 40 y 70 autores, y posteriormente para los artículos con entre 300 y 400 autores.
Luego, observamos que la tasa anual de citaciones tiene su moda alrededor de 1 citación anual, con valores mínimos cercanos a 1 citación cada 10 años y máximos cercanos a 100 citaciones por año.

Finalmente, confirmamos que existe una correlación de Pearson de 0.78 positiva entre el número de autores y la tasa anual de citaciones. 
Aunque los datos presentan un alto grado de dispersión, estos resultados respaldan hallazgos anteriores que indican que los artículos con un mayor número de autores tienden a acumular un mayor número de citaciones en un período determinado \citep{factores}.

\subsection*{Colombia en el contexto mundial}

Para ubicar a Colombia en el contexto mundial utilizamos los resultados de WoS para el periodo de cinco años entre el 2019 y el 2023.
Obtenemos el número total de publicaciones y el número de artículos que se consideran como altamente citados. 

La lista de naciones que seleccionamos es la unión de dos conjuntos. 
El primero corresponde a la lista de naciones reportadas por WoS con un número mayor o igual de publicaciones que Colombia. Estas son (en orden decreciente de publicaciones totales):
 Estados Unidos (USA),  Alemania (DEU), Inglaterra (ENG), Italia (ITA), China (CHN), Francia (FRA), España (ESP), Japón (JPN), Australia (AUS), Países Bajos (NLD), Chile (CHL), Canadá (CAN), Rusia (RUS), Suiza (CHE), India (IND), Suecia (SWE), Dinamarca (DNK), Corea del Sur (KOR), Polonia (POL), Escocia (SCT), Sudáfrica (ZAF), Brasil (BRA), Bélgica (BEL), México (MEX), Taiwan (TWN), Israel (ISR), Argentina (ARG), República Checa (CZE), Austria (AUT), Finlandia (FIN), Portugal (PRT), Hungría (HUN), Irlanda (IRL), Grecia (GRC), Irlanda del Norte (NIR), Gales (WLS), Nueva Zelanda (NZL), Emiratos Árabes Unidos (ARE), Ucrania (UKR), Noruega (NOR), Turquía (TUR), Irán (IRN), Tailandia (THA), Serbia (SRB), Eslovaquia (SVK), Croacia (HRV), Eslovenia (SVN), Bulgaria (BGR), Armenia (ARM), Estonia (EST) y Vaticano (VAT).
 El segundo conjunto de naciones corresponde a la lista de naciones de América Latina y el Caribe que no se encuentren en el conjunto anterior y que tengan al menos una publicación.
 Estas son:
 Uruguay (URY), Ecuador (ECU), Venezuela (VEN), Costa Rica (CRI), Perú (PER), Guatemala (GTM), Cuba (CUB), Honduras (HND), Bolivia (BOL), Nicaragua (NIC), El Salvador (SLV) y Paraguay (PRY)..

\begin{figure}[H]
    \centering
    \includegraphics[width=0.9\linewidth]{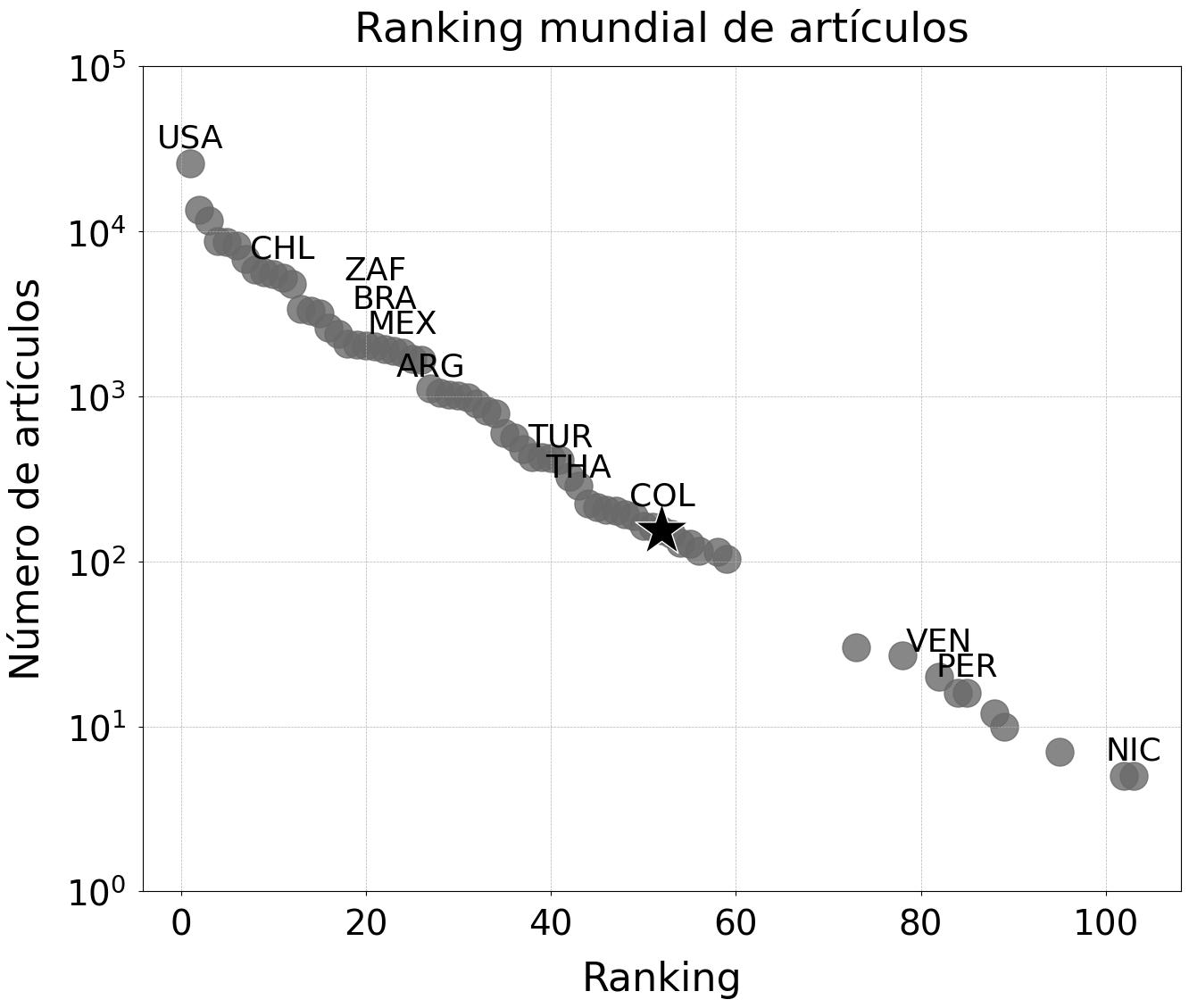}
    \caption{Distribución tamaño-rango para el número total de publicaciones 2019-2023 en WoS. Se incluyen las naciones que tienen un número igual o mayor de publicaciones que Colombia (COL), además de otras naciones de América Latina que tienen al menos una publicación. Colombia se encuentra en la posición 52 a nivel mundial.}
    \label{fig:ranking_mundial}
\end{figure}

La Figura \ref{fig:ranking_mundial} muestra el número total de publicaciones de todas estas naciones. Observamos que estos datos evidencian una relación exponencial, lo que indica que una distribución de ley de potencias no sería un buen ajuste, como se esperaría a partir de la amplia variedad de distribuciones tamaño-rango descritas por la ley de Zipf \citep{2009SIAMR..51..661C}.

En este ordenamiento por citaciones, los tres primeros lugares los ocupan Estados Unidos, Alemania e Inglaterra. Colombia se encuentra en el lugar 52 a nivel global. 
A nivel de América Latina, Colombia ocupa la quinta posición, después de Chile (11), Brasil (22), México (24) y Argentina (27), donde el número entre paréntesis corresponde a la posición global.
Dentro de América Latina y el Caribe, en posiciones inferiores del Ranking tenemos a Uruguay (73), Ecuador (78), Venezuela (82), Costa Rica (84), Peru  (85), Guatemala (88), Cuba (89), Honduras (95), Bolivia (102), Nicaragua (103), El Salvador (126) y Paraguay (130).

\begin{figure}[H]
    \centering
    \includegraphics[width=0.9\linewidth]{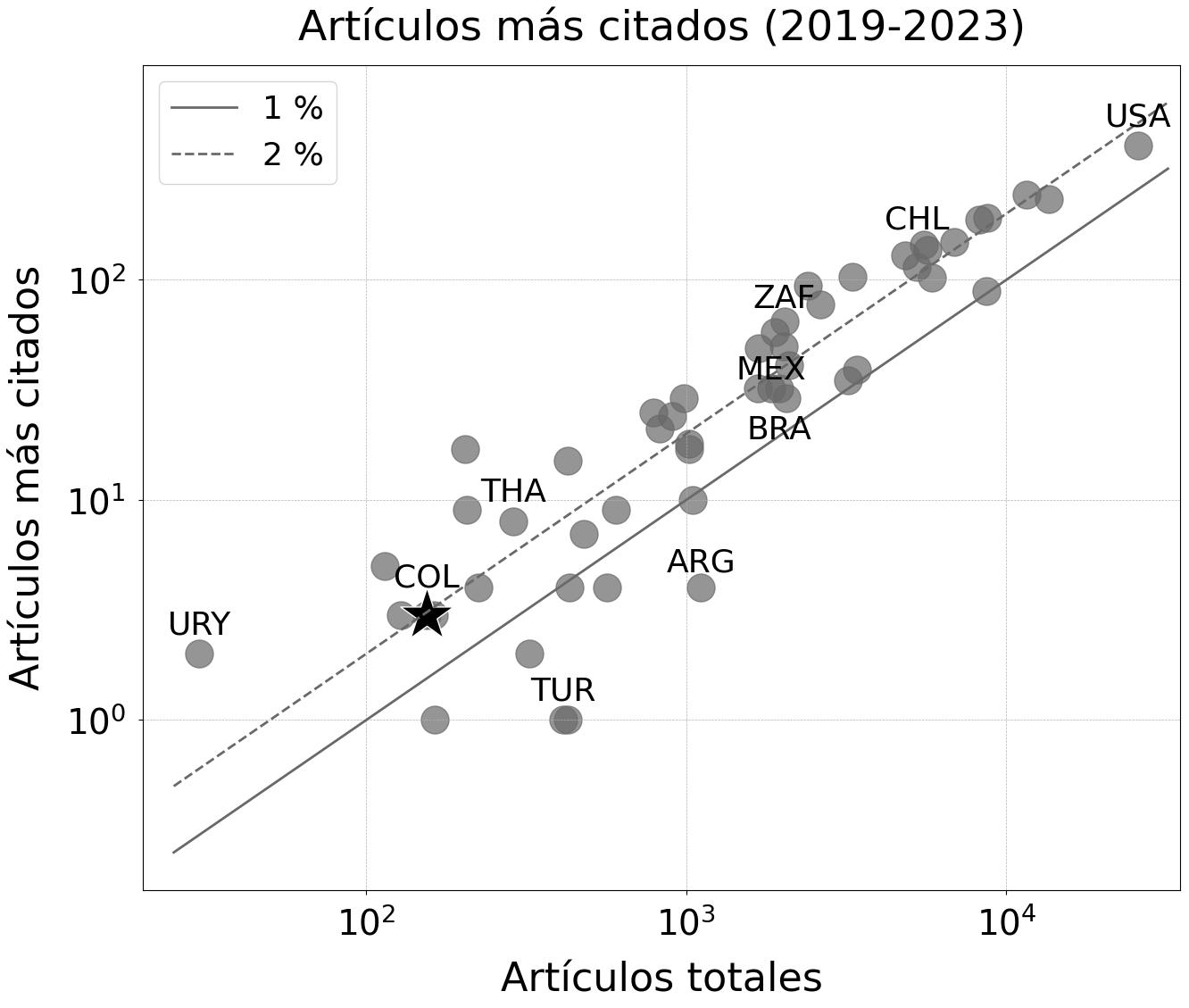}
    \caption{Artículos más citados (top 1\% a nivel mundial) entre el 2019 y el 2023 comparados con el número de artículos del 2019 a 2023, según WoS. 
    Mostramos además el 1\%  (línea continua) y el 2\% (línea discontinua) de los artículos totales de un país. 
    Colombia cuenta con 3 artículos en el top más citado y este representa aproximadamente el 2\% de todos los artículos publicados con participación colombiana.}
    \label{fig:altamente_citados}
\end{figure}

En la Figura \ref{fig:altamente_citados} mostramos el número de artículos altamente citados (en el top 1\% de citaciones a nivel global en su año de publicación) en función del número de artículos totales publicados. Es reconfortante ver que Colombia cuenta con 3 artículos en esta selección. Este número corresponde aproximadamente al 2\% de los artículos publicados con participación colombiana, una fracción comparable a la de otros países como Brasil, Chile y Estados Unidos.

Los tres artículos altamente citados fueron publicados en 2023 y corresponden a una colaboración de cosmología observacional llamada Dark Energy Spectroscopic Instrument:

\begin{itemize}
\item The Target-selection Pipeline for the Dark Energy Spectroscopic Instrument, \citep{2023AJ....165...50M}, con participación de Uniandes.
\item The Spectroscopic Data Processing Pipeline for the Dark Energy Spectroscopic Instrument, \citep{2023AJ....165..144G}, con participación de Uniandes.
\item The DESI Bright Galaxy Survey: Final Target Selection, Design, and Validation, \citep{2023AJ....165..253H}, con participación de Uniandes y la Universidad Tecnológica de Bolívar.
\end{itemize}

\section*{Conclusiones}
La investigación astronómica en Colombia está experimentando un periodo de crecimiento y consolidación. 
Sin embargo, los registros históricos de esta evolución no ofrecen una visión completa a nivel nacional \citep{Greiff,Cepeda,2017RMxAC..49....3H}. 
Este artículo busca llenar ese vacío mediante el análisis de publicaciones arbitradas con participación de instituciones colombianas, dada la importancia de este tipo de producción bibliográfica en la difusión del conocimiento y la reputación internacional de la comunidad astronómica del país.

Los datos y estadísticas presentados se basan en las publicaciones arbitradas encontradas en bases de datos como el Astrophysis Data System y Web of Science. 
Esta aproximación bibliométrica permite minimizar sesgos al no realizar una selección previa de las instituciones a estudiar. 
Por el contrario, el análisis de estas publicaciones nos permite identificar las principales universidades y áreas que han contribuido al crecimiento de la investigación astronómica en Colombia.

Esta perspectiva bibliométrica nos proporciona una serie de conclusiones significativas:

\begin{enumerate}

\item \textbf{Diversidad.} Encontramos que al menos 25 instituciones diferentes han participado en las publicaciones arbitradas que estudiamos con ADS. 
Estas instituciones están ubicadas en once ciudades diferentes, todas ellas pertenecientes a capitales departamentales (en paréntesis indicamos el número de instituciones): Barranquilla (1), Bogotá (11), Bucaramanga (1), Cali (2), Cartagena (1), Medellín (3), Montería (1), Pasto (2), Pereira (1), Popayán (1) y Villavicencio (1). 
Además la mayoría de estas instituciones (14 de 25) son de origen privado, mientras que la minoría (11 de 25) son de origen estatal.
Identificamos cuatro instituciones que tienen una clara preeminencia: la Universidad de los Andes, la Universidad Nacional de Colombia, la Universidad Industrial de Santander y la Universidad de Antioquia, distribuidas en tres ciudades diferentes.

\item \textbf{Los inicios.} El año 1980 marca de manera apropiada el surgimiento de la era contemporánea de la investigación astronómica en Colombia, evidenciado por la primera publicación en una revista internacional arbitrada de astronomía realizada por un científico afiliado a una institución colombiana.

\item \textbf{Cuatro décadas de evolución.} Durante la primera década (que denominamos \emph{fase pionera}), desde 1980 hasta 1989, las publicaciones arbitradas eran esporádicas con una tasa promedio de 1 publicación por año. 
Estas contribuciones estuvieron dominadas principalmente por dos instituciones en Bogotá: la Universidad Nacional de Colombia y la Universidad de los Andes. 
En las dos décadas subsiguientes (que denominamos \emph{fase de desarrollo}) la productividad se duplicó con un promedio de 2 publicaciones por año, desde 1990 hasta 2009.
Además dos instituciones más se desarrollaron en el ámbito de la investigación astronómica: la Universidad de Antioquia y la Universidad Industrial de Santander. 
En la década de 2010 hasta 2019 (que denominamos \emph{fase de consolidación}) la productividad se multiplica por un factor diez con respecto a la década anterior, con cerca de 20 publicaciones por año.

\item \textbf{El comienzo de la quinta década.} 
En los últimos cuatro años, de 2020 a 2023, se ha observado un nuevo crecimiento en la tasa de publicaciones, con cerca de 40 publicaciones por año. 
Durante este período, el número de publicaciones es comparable al total de las cuatro décadas anteriores (1980-2019). 
Esto se explica en parte por la participación dentro de colaboraciones científicas internacionales.

\item \textbf{Alto impacto.} Encontramos que los artículos con mayor número de citaciones abarcan tres áreas: fenómenos de altas energías y física fundamental, estrellas y física estelar, galaxias y cosmología. 

Uniandes muestra una concentración en galaxias y cosmología, con colaboraciones entre 7 y 32 autores, todos superando las 100 citaciones. 
En la UIS, se destacan las altas energías y la física fundamental, con colaboraciones numerosas (cerca de 400 autores), y la mayoría de las publicaciones superando las 100 citaciones. 
Para la UdeA, el enfoque está en estrellas y física estelar, con colaboraciones de una decena de autores y algunos artículos superando las 100 citaciones. 
En la UNAL, se abordan temas de estrellas y física estelar, galaxias y cosmología, con un grupo de cuatro autores por publicación y una media de citaciones menor que los artículos de Uniandes, la UIS y UdeA.

\item \textbf{Contexto internacional.} 
Para ubicar al país en el contexto mundial examinamos el total de publicaciones y los artículos altamente citados, enfocándonos en los resultados entre 2019 y 2023. 
La lista de países incluye aquellos con un número igual o superior de publicaciones que Colombia. 
En el número total de publicaciones destacan Estados Unidos, Alemania e Inglaterra como los líderes en publicaciones. 
Colombia ocupa el puesto 52 a nivel mundial y el quinto en América Latina. 
Tres artículos con participación colombiana están entre los más citados, todos de 2023 y relacionados con el Dark Energy Spectroscopic Instrument, una colaboración de cosmología observacional. 

\end{enumerate}

Esperamos que este trabajo pueda servir como un estímulo y una base sólida para futuras investigaciones. 
Por ejemplo, sería enriquecedor profundizar en el análisis de la colaboración internacional en proyectos astronómicos y su impacto en la producción científica de Colombia. 
Además, sería valioso investigar cómo las políticas gubernamentales y las estrategias de financiamiento pueden afectar el crecimiento y la calidad de la investigación astronómica en el país. 
Una línea de investigación prometedora sería explorar el papel de las instituciones educativas y los centros de investigación en la formación de nuevos talentos en astronomía y su contribución al desarrollo continuo del campo en Colombia. 
Asimismo, sería interesante explorar el potencial de la divulgación científica y la participación pública en actividades astronómicas para fomentar el interés y la comprensión pública de la ciencia en el país. 
Estas áreas ofrecen un terreno fértil para futuros investigadores interesados en impulsar y documentar el avance de la astronomía colombiana.

Nuestros resultados muestran que la investigación astronómica en Colombia, a diferencia de varios lugares en el mundo, se encuentra concentrada casi exclusivamente dentro de las universidades.
Esta situación presenta una ventaja significativa: es posible encontrar múltiples oportunidades de colaboración e intercambio interuniversitario que beneficien a todas las personas que realizan investigación astronómica en el país. 
Un primer paso hacia esta colaboración podría ser la creación de una alianza entre las universidades que realizan investigación en astronomía. 
De esta manera, se podrían optimizar los limitados recursos humanos y económicos disponibles en nuestras instituciones de educación superior en Colombia.

\section*{Agradecimientos}

Los autores expresan un sincero agradecimiento a Santiago Vargas,  Nelson Padilla y Mario Armando Higuera, cuyos comentarios y sugerencias contribuyeron significativamente a mejorar la versión original de este escrito.

\section*{Contribución de los autores}

SGM: recolección de datos, integración  de  datos,  cálculos, análisis  de  los  resultados, redacción del artículo
FOF: recolección de datos, integración  de  datos,  cálculos, análisis  de  los  resultados, redacción del artículo.
MPSA: revisión  de  la  literatura, redacción del artículo
PN: conceptualización, recolección de datos, revisión  de  la  literatura
JEFR: conceptualización, recolección de datos, revisión  de  la  literatura, análisis  de  los  resultados, redacción del artículo.
Todos los autores leyeron y aprobaron la versión final sometida a publicación.

\section*{Conflicto de intereses}

Los autores declaran que no tienen ningún conflicto de intereses que ponga en riesgo la validez de los resultados presentados.

\bibliographystyle{apacite}

\end{document}